\documentclass{ws-ijmpb}

\begin{document}

\markboth{RICHARD KERNER}
{Stochastic agglomeration and growth}

\catchline{}{}{}

\title{STOCHASTIC DESCRIPTION OF AGGLOMERATION AND GROWTH PROCESSES
IN GLASSES}

\author{\footnotesize RICHARD KERNER\footnote{Email: rk@ccr.jussieu.fr}}

\address{Laboratoire de Physique Th\'eorique des Liquides, Universit\'e
Paris VI - CNRS URA 7600, \\
Tour 22, $4^{eme}$, Boite 142, 4 Place Jussieu, Paris, France }

\maketitle

\pub{Received (received date)}{Revised (revised date)}

\begin{abstract}
We show how growth by agglomeration can be described by means
of algebraic or differential equations which determine the evolution
of probabilities of various local configurations. The
{\it minimal fluctuation} condition is used to define vitrification. 
Our methods have been successfully used for the description of glass 
formation.
\end{abstract}
\section{ Introduction } 
\label{sec:introduction}

\indent
In a series of papers published during the past ten years (\cite{Kerner95},
\cite{RKMM97a}, \cite{Kerner98a}, \cite{Kerner98b}), 
new models of growth by agglomeration of smaller units have been 
elaborated, and  applied to many important physical systems, 
such as quasicrystals (\cite{DMDSLRK94}), fullerenes
(\cite{RKKHBKP92}, \cite{Kerner94}), and oxide and chalcogenide glasses, 
(\cite{DMDSRKMM95}, \cite{RABJPDRK95},\cite{RBRKMMGGN97}, \cite{RKGGN98}).
Here we shall present the main ideas on
which these models are based, and briefly discuss the latest developments.
\newline
\indent
In order to make our presentation concise, the example 
we choose is the simplest covalent network glass known to physicists, 
the binary chalcogenide glass $As_x Se_{(1-x)}$, where $x$ is the 
concentration of arsenic atoms in the basic glass-former, which in this 
case is pure selenium. The generalization to other covalent networks,
e.g. $Ge_x Se_{(1-x)}$, is quite straightforward. These glasses 
(in the form of thin and elastic foils) are used in photocopying devices.
\newline
\indent
Whether the formation of a solid network of atoms or molecules occurs in a
more or less rapidly cooled liquid, or as vapor condensation on a cold support, 
the most important common feature of these processes is progressive 
agglomeration of small and mobile units (which may be just single atoms, 
or stable molecules, or even small clusters already present in the liquid state)
into an infinite stable network, whose topology can no longer be modified 
unless the temperature is raised again, leading to the inverse (melting or 
evaporation) process.
\newline
\indent
To describe such an agglomeration with all geometrical and physical parameters,
such as bond angles and lengths, and the corresponding chemical and mechanical
energies stored in each newly formed bond, is beyond the possibilities of any
reasonable model. This is why stochastic theory is an ideal tool
for the description of random agglomeration and growth processes. Instead of
reconstructing all local configurations, it takes into account only the
{\it probabilities} of them being found in the network, and then the
probabilities of higher order, corresponding to local correlations. This is
achieved by using the {\it stochastic matrix} technique. A {\it stochastic
matrix} \, {\bf M} \, represents an operator transforming given finite
distribution of probabilities , $[p_1, p_2, ..., p_N]$ , into another 
distribution of probabilities, $[p^{'}_1 , p^{'}_2 , ..., p^{'}_N]$.
It follows immediately that such a matrix must have only real non-negative
entries, each column summing up to $1$.
\newline
\indent
The algebraic properties of such matrices are very well known. The main
feature that we shall use here is the fact that any stochastic matrix has
at least one eigenvalue equal to $1$. The remaining eigenvalues have
their absolute values always less than $1$. This means that if we continue
to apply a stochastic matrix to any initial probability distribution, after
some time only the distribition corresponding to the unit eigenvalue will
remain, all other contributions shrinking exponentially. This enables us to
find the asymptotic probability distribution.

In what follows, we identify these probability distributions with
stable or meta-stable states of the system, fixing the statistics of
characteristic sites in the network. Taking into account Boltzmann factors
(with chemical potentials responsible for the formation of bonds), we are
able to find the glass transition temperature in various compounds. In 
particular, one is able to predict the initial slope of the curve $T_g(c),$  
i.e. the value of $(dT_g/dc )_{c=0}$ (\cite{Kerner94a}, \cite{RKMM97}).
\section{Stochastic matrix describing cluster agglomeration}
\label{sec:stochastic}
Consider a binary selenium-arsenic glass, in which selenium is the basis
glass former, and arsenic is added as modifier (although its concentration
can be as high as $30\%$). The chemical formula denoting this compound
is $As_c Se_{(1-c)}$, where $c$ is the $As$
concentration. In a hot liquid, prior to solidification, the basic
building blocks that agglomerate are just selenium and arsenic atoms,
indicated respectively by
$\left(\mbox{---\hspace{-1pt}$\circ$\hspace{-1pt}{---}}\right)$ and
$\left(\mbox{---\hspace{-4pt}$\bullet$\hspace{-4pt}\raisebox{-4pt}
{$\backslash$}\hspace{-5pt}\raisebox{4pt}{/}}\right)$. When the temperature
goes down, clusters of atoms start to appear everywhere, growing by
agglomeration of new atoms on their rim. Consider a growing cluster:
one can distinguish three types of situations (we shall call them
{\it ``sites''}) on the cluster's rim. The concentration of free $As$
atoms in the liquid will be called $c$ and that of $Se$, $(1 - c)$.

We should stress here that the mathematical model we propose to analyze is
quite far from reality in the particular case of $As-Se$ binary glass, as
has been shown in (\cite{Georgiev2000}), because some of the $AS$ atoms
are five-coordinated. Our model gives better predictions for the
$Ge-Se$ binary glasses. Our aim here is to expose the basic theoretical
concepts rather than get very precise predictions; for this, one needs more
sophisticated models, taking into account various possibilities of different
agglomeration modes, e.g. ring-forming and coordination changes.

Two choices are possible for constructing the states and transition matrix 
(see (\cite{RARK1})). There are three possible kinds of sites: a selenium
atom with one unsaturated bond, and an $As$ atom presenting one or two 
free bonds; these are indicated by $x$ = \mbox{$\circ$\hspace{-1pt}{---}}, $y$ =
\mbox{$\bullet$\hspace{-4pt}\raisebox{-4pt}
{$\backslash$}\hspace{-5pt}\raisebox{4pt}{/}} and $z$ =
\mbox{$\bullet$\hspace{-1pt}{---}}. To each site one of the two basic
cells can attach itself, reproducing one of the initial configurations,
in the specific combinations shown in the next column of the Figure 1. The
attachment of {\em one single basic cell}, or the saturation of one
single bond, is a step in the evolution. In the second choice, each
step is obtained by the complete saturation of all the bonds at the rim,
so that only two types of sites (denoted by $x$ and $y$) are seen on
cluster's rim, assuming that the growth is of dendritic type 
(no small rings present). 
It can be shown (\cite{RARK1}) that the two approaches lead to the same
results, which may be considered as a proof of the ergodicity of the
proposed model. We shall choose the second version of the model for the
sake of simplicity. In this case, we can take into account only the $x$
and $y$-type sites, because the $z$-type sites transform after the next 
agglomeration step into an $x$ or $y$ type site. The elementary step in 
the agglomeration process, described by  he transition matrix, corresponds 
now to the complete saturation of all  the available free bonds on the rim. 
This is represented in Figure 1 :

\begin{center}

    \begin{tabular}{|c c c c c}

 &  \hspace{50pt} &
 \hspace{18pt}{{\mbox{---\hspace{-1pt}$\circ$\hspace{-1pt}{---}}
 \hspace{-4pt}$\circ$\hspace{-1pt}{---}}}  & {\large x}
 & \hspace{50pt} 2 (1 - c) $e^{- \epsilon}$
 \\

\hspace{-6pt}{\mbox{---\hspace{-1pt}$\circ$\hspace{-1pt}{---}}}
& {\large x} & &
\\
 & \hspace{50pt} &
 \hspace{12pt}{{\mbox{---\hspace{-1pt}$\circ$\hspace{-1pt}{---}}
\hspace{-8pt} $\bullet$\hspace{-3pt}\raisebox{-5pt}
    {$\backslash$}\hspace{-5pt}\raisebox{5pt}{/}} } &
{\large y} & \hspace{50pt} 3 c $e^{- \eta}$                                                                                                                          %
\\
\\

 &  \hspace{50pt} & \hspace{6pt}\mbox{{\bf ---}
\hspace{-4pt}$\bullet$
\hspace{-9pt}\raisebox{-1pt}{{\Large$<$}}
\hspace{-6pt}{\raisebox{4pt}{$\circ$
\hspace{-4pt}{\bf ---}}}}
\hspace{-20pt}{\raisebox{-4pt}{$\circ$
\hspace{-2pt}{\bf ---}}} & {\large 2 x}
& \hspace{50pt} 4 $(1 - c)^{2}$ $e^{- 2 \eta}$

\\

\hspace{-13pt}{--}\hspace{-5pt}
\mbox{---\hspace{-5pt}$\bullet$\hspace{-4pt}\raisebox{-5pt}
{$\backslash$}\hspace{-6pt}\raisebox{5pt}{/}} & {\large y} &
\hspace{6pt}\mbox{{\bf ---}\hspace{-2pt}$\bullet$
\hspace{-9pt}\raisebox{-1pt}{{\Large$<$}}\hspace{-5.5pt}
{\raisebox{4pt}{$\circ$ \hspace{-1pt}{\bf ---}}}\raisebox{-4pt}
{\hspace{-19pt}$\bullet$ \hspace{-9pt}\raisebox{-1pt}{{\Large$<$}}}}
& {\large x + y} &
\hspace{50pt} 12 c (1 - c) $e^{- \eta - \alpha}$
\\
                                                                                                                                                                    
 &  \hspace{50pt} &
\mbox{---\hspace{-3pt}$\bullet$\hspace{-5pt}\raisebox{-5pt}
{$\backslash$}\hspace{-5pt}\raisebox{5pt}{/}}
 \mbox{\hspace{-6pt}\raisebox{-10pt}{$\bullet$
\hspace{-6pt}\raisebox{-1pt}{{\Large$<$}}}
\raisebox{10pt}{{\hspace{-14pt}$\bullet$
\hspace{-8pt}\raisebox{-1pt}{{\Large$<$}}}}}
& {\large 2 y}
& \hspace{50pt} 9 $c^{2}$ $e^{- 2 \alpha}$

    \\

    \end{tabular}

   \ 

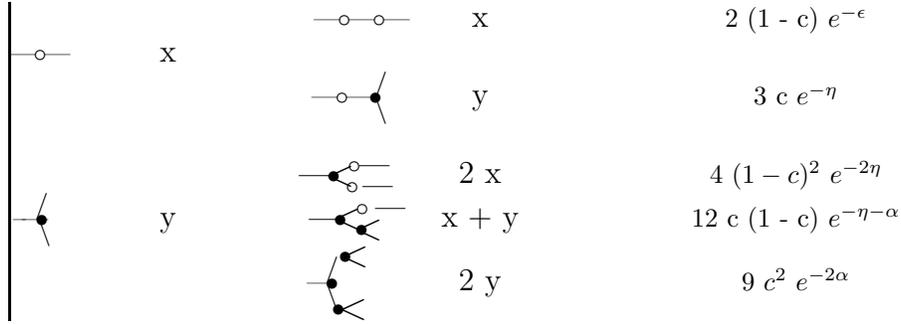
Figure 1: {\it States, steps and un-normalized probability factors .}
\end{center}

Observing that from the site $z$ only the sites of $x$ and $y$ type can be
produced, we can forget it and consider the dendritic growth with only two
types of sites appearing all the time. Given an arbitrary initial state
$(p_x ,  p_y)$, the new state results from taking into account all possible
ways of saturating the bonds of the previous state's sites by the available
external atoms.  The un-normalized probability factors are displayed in the
Figure. The non-normalized probability factors can be arranged in a matrix
\begin{equation}
\pmatrix{ 2 (1-c) e^{-\epsilon} & 4 (1-c)^2
e^{-2 \eta} \cr 8 (1 - c)^2 e^{- 2 \eta} +
12 c (1-c) e^{- \eta - \alpha} &
12 c (1-c) e^{- \eta - \alpha} + 18 c^2
e^{- 2 \alpha} }
\label{unnorm}
\end{equation}
The normalized transition matrix is written as 
\begin{equation}
M = \pmatrix{M_{xx} & M_{xy} \cr M_{yx} & M_{yy}} =
\left(
\begin{array}{cc}
M_{xx} & 1- M_{yy} \\
1 - M_{xx} & M_{yy}
\end{array} \right) 
\label{rimM}
\end{equation}
\noindent
where the entries are obtained by normalizing the columns of
the matrix (\ref{unnorm}).
\begin{equation}
M_{xx} = \frac{2(1-c) \xi}{2(1-c) \xi + 3 c}\,  , \, \ \ {\rm and} \, \ \
\, \ \ M_{yy} = \frac{3 c \mu}{2 (1-c) + 3 c \mu}
\label{defAB}
\end{equation}
where we have introduced the abbreviated notation $\xi = e^{\eta -
\epsilon}$ and $\mu = e^{\eta - \alpha}$.
\newline
\indent
The eigenvalues of this matrix are
$1$ and $M_{xx} - M_{yy} = M_{xy}-M{yx}$.  
and the stationary eigenvector is
\begin{equation}
\pmatrix{p_x^{\infty} \cr p_y^{\infty}} = \frac{1}{M_{xy}+M_{yx}} \left(
\begin{array}{c}
M_{xy}\\ M_{yx}
\end{array}
 \right) \; ,
\end{equation}
\indent
It can be seen from  Figure 1 that on the surface of an average cluster,
$p_{x}$ is the $Se$ concentration and $p_{y}$ is the $As$ concentration.
Now, the high homogeneity exhibited by known glass structures suggests
that even in relatively small clusters, deviations from the
average modifier concentration $c$ must be negligible.  Thus, {\em in
the bulk}, the $As$ concentration should be equal to $c$. Therefore, the
condition of minimal fluctuations in the bulk concentration can be 
interpreted  as the glass transition condition. This means that 
the asymptotic state is fixed by the external concentration, therefore
the above eigenvector must be equal to the average distribution vector
$\left(1 - c, c \right) $.  The solutions are $c = 0$, $c = 1$ and the
nontrivial one %
\begin{equation}
c  = \frac{M_{yx}}{M_{xy}+M_{yx}} = \frac{6 - 4 \xi}{12 - 4 \xi  - 9 \mu } \; . 
\label{goodcondition}
\end{equation}
This equation can be checked against experiment.  For example, we
can evaluate the derivative $\frac{\partial T}{\partial c}$ =
$\left(\frac{\partial c}{\partial T}\right)^{-1}$ for a given value of
$c$.  In particular, as $c \rightarrow 0$, we can neglect the
$As$--$As$ bond creation (equivalent to putting $\mu = 0$ in
(\ref{goodcondition})), to get %
$$ %
\left[ \frac{\partial T}{\partial c} \right]_{c=0} =
\frac{T_{g_{0}}}{\ln (3/2)} \; , $$ %
 (where $T_{g_{0}}$ is the glass transition temperature of pure $Se$).
 This is the present--case expression of the general formula given by
 the stochastic approach, the fraction $(3/2)$ being replaced by 
$(m^{\prime}/m)$, where $m$ and $m^{\prime}$ are the valences of the basic 
glass former and of the modifier), remaining in very good agreement with the
experimental data (see \cite{Boolchand97}, \cite{Boolchand2000}, 
\cite{Georgiev2000}).

\section{Low concentration limit.}
\label{Low limit}
The above scheme can be easily generalized to the case of arbitrary valence,
say $m_A$ and $m_B$. In that case, the stochastic $2 \times 2$ matrix has the
same form as (\ref{rimM}), but with the entries given by
$$\, \ \ M_{xx} = 1 - M_{yx} = \frac{m_A (1-c) \xi}{m_A (1-c) \xi + m_B c} \, , \, \ \ \, 
\ \ M_{xy} = 1 - M_{yy} = \frac{m_A (1-c) }{m_A (1-c) + m_B c \mu} \, , $$
The asymptotic probability has the same form as before, as well as the zero
fluctuation condition relating $c$ with $T$ (interpreted as the glass
transition temperature). The derivative of $c$ with respect to the
temperature $T$ gives the ``magic formula''
\begin{equation}
\frac{d \, c}{d \, T} = \frac{1}{T} \, \frac{(\frac{m_A}{m_B} - \mu) \, \xi
ln \xi - (\frac{m_B}{m_A} - \xi) \, \mu ln \mu }{ [(1 - \frac{m_A}{m_B}
 \xi ) + ( 1 - \frac{m_B}{m_A} \mu ) ]^2 }
\label{derivative}
\end{equation}
where we used the fact that $ \frac{d \, \xi}{ d \, T} = - \frac{1}{T} \, \xi
\, ln \xi$ , \,  and \,  $ \frac{d \, \mu}{ d T} = - \frac{1}{T} \, \mu \,
ln \mu . $ This defines the slope of the function $T_g(c)$, which is
an important measurable quantity :
\begin{equation}
\frac{d \, T_g}{d \, c} = T_g \, \, \frac{[(1- \frac{m_A}{m_B} \,\xi )
+ (1 - \frac{m_B}{m_A} \, \mu) ]^2}{(\frac{m_A}{m_B} - \mu) \, \xi ln \xi
- (\frac{m_B}{m_A} - \xi) \, \mu ln \mu }
\label{derivative2}
\end{equation}
The initial slope, at $c =0$, is of particular interest. Its expression is
very simple, taking into account that when $c = 0$, we have also
$\xi = \frac{m_B}{m_A}$, which leads to
\begin{equation}
\biggl[ \frac{d \, T_g}{d \, c} \biggr]_{c=0} = \frac{T_{g0} \, (1 -
\frac{m_B}{m_A} \, \mu)}{ln ( \frac{m_B}{m_A})}
\end{equation}
\indent
Its value has been checked against the experiment very successfully, in
more than 30 different compounds. In some cases the formula does not
seem to work well; usually it comes from the change of valence of certain
atoms provoked by the influence of the surrounding substrate.

One could be worried about the apparent singularity in this
formula when $m_A = m_B$, i.e. when one deals with a mixture
of two different glass formers with the same coordination number.
It is not difficult to show that also in such a case a
reasonable limit can be defined, as has been recently suggested
by M.Micoulaut (\cite{MMprivate}). As a matter of fact, suppose that 
the glass transition temperature of the
pure glass-former $A$ is $T_{g0}$, and that of the pure
glass-former $B$ is $T_{g1}$. We can re-write our minimal
fluctuation condition in a very symmetric manner, invariant
with respect to the simultaneous substitution $m_A \leftrightarrow m_B$,
$c \leftrightarrow (1-c)$ and $\xi \leftrightarrow \mu$ \, :
\begin{equation}
c(1-c) \, [ (1-c) \, (1 - \frac{m_A}{m_B} \, \xi ) -
c \, ( 1 - \frac{m_B}{m_A} \, \mu  ) ] = 0
\label{symcondition1}
\end{equation}
Obviously, the ``pure states'' $c = 0$ or $c = 1$ represent stationary
solutions of (\ref{symcondition1}) and can be factorized out. The non-trivial
condition for the glass forming is thus
\begin{equation}
(1-c) \, [ 1 - \frac{m_A}{m_B} \, \xi ] - c \, [
1 - \frac{m_B}{m_A} \, \mu ]  = 0
\label{symcondition2}
\end{equation}
Now, using the limit conditions at $c \rightarrow 0,  \, \ \ T_g = T_{g0}$
and $c \rightarrow 1 , \,  \ \ T_g = T_{g1}$, and
introducing the generalized Boltzmann factors with the energy barriers
for corresponding bond creations as $E_{AA}, \, E_{AB} \,$ and $E_{BB}$,
we can write
\begin{equation}
E_{AB} - E_{AA} = k \, T_{g0} \, ln ( \frac{m_B}{m_A}) \, ,
\, \ \ \,
E_{AB} - E_{BB} = k \, T_{g1} \, ln ( \frac{m_A}{m_B} ) , 
\label{energies3}
\end{equation}
so that the expressions $\xi$ and $\mu$ at the arbitrary temperature $T$
can be written as
\begin{equation}
\xi (T) = e^{\frac{E_{AB} - E_{AA}}{ T_{g0}}  \cdot \frac{T_{g0}}{T}}
= ( \frac{m_B}{m_A} )^{\frac{T_{g0}}{T}} \, ; \, \ \ 
\mu  (T) = e^{\frac{E_{AB} - E_{BB}}{ T_{g1}}  \cdot \frac{T_{g1}}{T}}
= ( \frac{m_A}{m_B} )^{\frac{T_{g1}}{T}} . 
\label{xi}
\end{equation}
Substituting these expressions into  (\ref{derivative2}) and
taking the limit $c \rightarrow 0$, we get
\begin{equation}
\frac{d T_g}{dc}\mid_{c=0} = \frac{ T_{g0} [1 - (\frac{m_B}{m_A})^{\frac{ T_{g0} -
T_{g1}}{T_{g0}}} ] }{ln(\frac{m_B}{m_A})} 
\end{equation}
It is easy to see now that even when $m_A = m_B$, this formula has
a well defined limit. Indeed, if we first set $\frac{m_B}{m_A} =
1 + \epsilon $, and then develop the numerator and the denominator of the
above equation in powers of $\epsilon$, then in the limit when $\epsilon
\rightarrow 0 $ we arrive at a simple linear dependence which is
in agreement with common sense and with experiment as well, namely
\begin{equation}
\frac{d T_g}{dc} \mid_{c=0} = T_{g1} - T_{g0}
\label{Tlinear}
\end{equation}
\indent
This formula is also confirmed by many experiments, e.g. performed on 
selenium-sulfur mixtures (where $m_A = m_B = 2$). The deviations from 
the linear law (\ref{Tlinear}) observed in the $Se-Te$ binary glass are
explained by the fact of the chemical properties of tellurium, which
changes its valence from $2$ to $3$ in presence of selenium.
\section{The effect of rapid cooling}
\label{cooling}
\indent
An interesting extension of this model is obtained when we take into account
the effects of rapid cooling, i.e. when the time derivative of the temperature
can no longer be neglected. The treatment of this problem was suggested
in (\cite{RKMM1992}), and has been solved quite recently (\cite{Kerner2002}).
\newline
\indent
Consider the agglomeration process defined by  the above stochastic matrix,
${\vec{p}}^{\, \prime} = M \vec{p}$, with $\vec{p}$ representing a normalized
column (a ``vector'') with two entries, $p_x$ and $p_y = 1 - p_x$. After one
agglomeration step, representing on the average one new layer formed on the
rim of a cluster, we can write
\begin{equation}
\Delta \vec{p} = {\vec{p}}^{\, \prime} - \vec{p} = (M - {\bf 1}) \, \vec{p}
\label{deltap}
\end{equation}
Let us introduce a symbolic variable $s$ defining the progress of the
agglomeration process; obviously, $s(t)$ should be a monotonically increasing
function during the glass transition. If the temperature variation is so
slow that the derivative $dT/dt = (dT/ds)(ds/dt)$ can be neglected (which
is often called the {\it annealing} of glass), the master equation of our
model can be written as
$$ \Delta \vec{p} = \frac{\partial \vec{p}}{\partial s} \, \Delta s =
(M - {\bf 1}) \, \vec{p} \, \Delta s $$
where the variation $\Delta s$ represents one complete agglomeration step.
If we want to use real time $t$ as an independent parameter, we should write
\begin{equation}
\frac {d \vec{p}}{dt}= \frac{\Delta \vec{p}}{\Delta s} \frac{d s}{dt} =
\tau^{-1} \, \frac{\Delta \vec{p}}{\Delta s} = \frac{1}{\tau} \, (M - {\bf 1})
\, \vec{p}
\label{stepder}
\end{equation}
We have introduced here the new entity $\tau = (ds/dt)^{-1}$ which can be
interpreted as the average time needed to complete a new layer in any
cluster, or alternatively, the time needed for an average bond creation.
Now, if the temperature varies rapidly enough, the matrix $M$ can no longer be
considered as constant. The equation ({\ref{stepder}) must be
modified according to the well known  ``moving target'' principle. That is,
the {\it total} derivative of $\vec{p}$ with respect to $t$ should read:
\begin{equation}
\frac{d \vec{p}}{dt} = ( M - {\bf 1}) \, \frac{ds}{dt} \, \vec{p} \, +
\frac{\partial M}{\partial T} \, \frac{d T}{d t} \, \vec{p} =
\frac{d \vec{p}}{dt} = \biggl[ \frac{1}{\tau} \, (M-{\bf 1}) \, + q \,
\frac{\partial M}{\partial T} \biggr] \, \vec{p}
\label{target1}
\end{equation}
where we supposed linear dependence of the temperature on time, so that
the derivative $dT/dt$ can be denoted by constant cooling rate $q$.
In the two-dimensional case only one component of $\vec{p}$ is independent,
because $p_x + p_y = 1$. Let us choose $p_y$ (whose asymptotic value
should be equal to $c$) as independent variable. Then (\ref{target1}) will
reduce to the single equation :
\begin{equation}
\frac{d p_y}{dt} =  \frac{1}{\tau} \, \biggl[ \, (M_{yy} - 1) \, p_y +
M_{yx} \, (1-p_y) \biggr] + q \, \biggl[ \frac{\partial M_{yy}}{\partial T}
\, p_y + \frac{\partial M_{yx}}{\partial T} \, (1-p_y) \biggr] 
\label{target3}
\end{equation}
where we have used the fact that $p_x = 1-p_y$ , $M_{xx} = 1 - M_{yx}$ and
$M_{yy} = 1 - M_{xy}$.
\newline
\indent
What remains is just simple algebra. After a few operations we find the
asymptotic value of $p_y$, denoted $p_y^{\infty}$, obtained when we set
$dp_y/dt = 0$ :
\begin{equation}
p_y^{\infty} =
\frac{M_{yx}+ \tau q \, (\frac{\partial M_{yx}}{\partial T})}{(M_{xy} +
M_{yx}) + \tau q (\frac{\partial (M_{xy} + M_{yx})}{\partial T})}
\label{q-equilibre}
\end{equation}
As in the former case, we define the glass transition temperature by solving
the zero-fluctuation condition $p_y^{\infty} = c$. The quasi-equilibrium
condition thus obtained can be written in a form displaying an apparent
symmetry between the two ingredients ("A" and "B") of binary glass.
As in the previous case (when $q = 0$), the limit values $c=0$ and $c=1$
represent stationary solutions, which is obvious (no local fluctuations of
concentration $c$ are possible when there is no ingredient other than $A$
or $B$ atoms alone). After factorizing out $c(1-c)$, we get
$$ \frac{m_B }{m_A (1-c) \xi + m_B c} - \frac{m_A }{m_A (1-c)+ m_B c \mu}= $$
\begin{equation}
\frac{\tau \, q}{T} \, m_A m_B \, \biggl[ \, \frac{c \, \mu \, ln
\mu}{[ m_A (1-c) + m_B c \mu]^2} -   \frac{(1-c) \, \xi \, ln \xi}{[ m_A
(1-c) \, \xi + m_B c ]^2} \biggr]
\label{symquench2}
\end{equation}
where we have used the fact that 
$\frac{\partial (ln \, \xi)}{\partial T} = - \frac{ln \, \xi}{T} \, , \, \ \ \, 
\, \ \  \frac{\partial (ln \,\mu)}{\partial T} = - \frac{ln \, \mu}{T} \, . $ 
\indent
The above formula seems quite cumbersome, but it become much simpler in the
low concentration limit, $c \rightarrow 0$ Close to $c=0$ we get
\begin{equation}
\frac{m_B}{m_A} - \xi + \frac{\tau \, q}{T} \, \frac{m_B}{m_A} \, ln \, \xi
= 0
\label{quenchc0}
\end{equation}
(quite obviously, in the limit $c \rightarrow 1$ one gets the same formula
switching $m_A$ with $m_B$ and replacing $\xi$ by $\mu$). Replacing 
$\xi$ by the expression (\ref{xi}), we arrive at :
\begin{equation}
\biggl[ \,1 - \biggl(\frac{m_B}{m_A}\biggr)^{\frac{T_{g0}-T}{T}} \, \biggr]+
(\frac{\tau \, q}{T}) \, \frac{T_{g0}}{T} \, ln \, (\frac{m_B}{m_A} )= 0 \, .
\label{quenchtemp}
\end{equation}
It is easy to see that independently of the ratio $m_B/m_A$, for temperatures
$T$ {\it above} $T_{g0}$ we must have $q < 0$, and vice-versa, during rapid
cooling the glass transition occurs at the temperature $T > T_{g0}$.
\newline
\indent
The dimensionless combination $(\tau \, q)/T$ defines the quenching
rate as the product of $(1/T)(dT/dt)= d(ln T)/dt$ by the
time constant $\tau$, characterizing the kinetics of the agglomeration process,
i.e. the average time it takes to create a new bond. It may depend
weakly on the temperature, but for the sake of simplicity suppose it is
constant. It can be determined by comparing formula (\ref{quenchtemp})
with the experimental data. To take an example, let us again consider the
selenium-arsenic glass at $c\rightarrow 0$ (almost pure selenium with a
small addition of $As$). We know that in this case $T_g \rightarrow T_{g0} =
318^0 K$. The formula (\ref{quenchtemp}) then gives the quasi-linear
dependence of $\Delta T = T - T_{g0}$ on the quenching rate $q$: for
$T_g= 328^0 K$ (i.e. $\Delta T = 10^0 K$) we get $\tau \, q = - 10.38 ;$
for $T_g=338^0 K$ (i.e. $\Delta T = 20^0 K$) we get $\tau \, q = - 21.51;$
for $T_g=348^0 K$ (i.e. $\Delta T = 30^0 K$) we get $\tau \, q = - 32.26,$
and so forth.
\newline
\indent
Finally, if we want to establish the formula for a pure glass-former, without
any modifier, we should take the limit $(m_A/m_B) \rightarrow 1$ and $\mu 
\rightarrow \xi$; we then get
\begin{equation}
\frac{T - T_{g0}}{T} + (\frac{\tau \, q}{T}) \, \frac{T_{g0}}{T} =  0 \, \ \ 
{\rm or} \, \ \ \, \ \ 
T - T_0 = \Delta T_g = - (\tau \, q) \, \frac{T_{g0}}{T} .
\end{equation}
\indent
Eventually, the deviations from this simple dependence may indicate that the
characteristic time $\tau$ depends on $T$. This can shed more light 
on the agglomeration kinetics in various glass-forming liquids.
More details can be found in (\cite{KernerStPbg}), (\cite{Kerner2002})

\section*{Acknowledgements}
Enlightening discussions with R.Aldrovandi, R.A.Barrio and M.Micoulaut are
gratefully acknowledged.


\begin{thebibliography}{0}

\bibitem{Kerner95}
R. Kerner, Physica {\bf B215} (1995) 267

\bibitem{RKMM97a}
R.Kerner and M.Micoulaut, Journ. of Physics: Cond. Matter, {\bf 9}, 
2551-2562, (1997).

\bibitem{Kerner98a}
R.Kerner, {\it A theory of glass formation}, in {\it Atomic diffusion in
amorphous solids}, M. Balkanski and R.J. Elliott, eds., World Scientific,
p.25-80  (1998)

\bibitem{Kerner98b}
R.Kerner, {\it The principle of self-similarity and its applications to
the description of non-crystalline matter,}  Proceedings of the Workshop 
in Cocoyoc, 1997, J.L. Moran-Lopez ed., 
Plenum Press, 323-337, (1998).

\bibitem{DMDSLRK94}
D.M. dos Santos-Loff and R.Kerner, Journal de Physique {\bf I} (4),
1491-1511, (1994

\bibitem{RKKHBKP92}
R. Kerner, K.H. Bennemann, K. Penson, Europhysics Lett. {\bf 19} (5), 363-368
(1992).

\bibitem{Kerner94}
R. Kerner, Computational Materials Science, {\bf 2} p.500-508, (1994); 

\bibitem{DMDSRKMM95}
D.M. dos-Santos-Loff, R.Kerner and M.Micoulaut, Journ. of Phys. C,
{\bf 7}, 8035-8052, (1995).

\bibitem{RABJPDRK95} 
R.A.Barrio, J.P.Duruisseau and R.Kerner,  
Phil. Magazine B, {\bf 72} (5), 535-550, (1995).

\bibitem{RBRKMMGGN97}
R.A.Barrio, R.Kerner, M.Micoulaut, G.G.Naumis, Journ. of.Phys.: Condensed
Matter, {\bf 9} p.9219-9234 (1997)

\bibitem{RKGGN98}
R.Kerner, G.G.Naumis, Journal of Non-Cryst.Solids {\bf 231} p.111-119 (1998)
and 
R,Kerner, G.G.Naumis, Journ. of Physics: Cond. Matter {\bf 12} (8),
p. 1641-1648, (2000).

\bibitem{Kerner94a}
R.Kerner, Journ. of Non-Cryst.Solids, {\bf 182}, p.9-21 (1995)

\bibitem{RKMM97}
R.Kerner, M.Micoulaut, Journ. of Non-Cryst. Solids, {\bf 210}, p.298-305 (1997)

\bibitem{RARK1}
R.Aldrovandi and R.Kerner, in {\it ``New Symmetries and Integrable Models''},
Proceedings of XIV Max Born Symposium, eds. A. Frydryszak, J. Lukierski and
Z. Popowicz, World Scientific, p. 153-169, (2000)

\bibitem{Boolchand97} 
P.Boolchand and W.J.Bresser, Phil.Mag.B {\bf 80}, 1757-1772 (2000)

\bibitem{Boolchand2000}
Boolchand P., Feng X. and Bresser W.J., {\it Rigidity transition in
binary $Ge-Se$ glasses}, Journ. Non-Cryst.Solids {\bf 293}, p.348 (2001)

\bibitem{Georgiev2000}
D.G.Georgiev, P.Boolchand and M.Micoulaut, Phys.Rev.B {\bf 62}, R9228 (2000)

\bibitem{KernerStPbg}
R.Kerner, Mathematical models of glass formation , Proceedings of the
conference "Glasses and Solid Electrolytes " (St.Petersburg, May 1999),
 Glass Physics and Chemistry, {\bf 26} (4), p.313 - 324 (2000)

\bibitem{MMprivate}
M.Micoulaut, private communication (2001)

\bibitem{RKMM1992}
R.Kerner, M.Micoulaut, C.R.Acad.Sci.Paris, t. {\bf 315}, S\'er.II, p.1307-1313
(1992)

\bibitem{Kerner2002}
R.Kerner and J.C.Phillips, Solid State Communications, {\bf 117}, p.47-56
(2000);  R.Kerner, in preparation. 

\end{thebibliography}
\end{document}